\documentclass[twocolumn,showpacs,amsmath,amssymb,prb,superscriptaddress,a4paper]{revtex4}

  \usepackage[T1]{fontenc}
  \usepackage[utf8]{inputenc}
  \usepackage[frenchb, english]{babel}
  \usepackage{amsmath}
  \usepackage{xcolor}
  \usepackage{graphicx}
  \usepackage{graphics}
  \usepackage{dcolumn}
  \usepackage{pgf,pgfarrows,pgfautomata,pgfheaps,pgfnodes,pgfshade}
  \usepackage{multimedia}
  \usepackage{braket}
  \usepackage{lmodern}
  \usepackage{picture}
  \usepackage{psfrag,fancyhdr,layout,appendix}
  \usepackage{epstopdf}
  \usepackage{placeins}
  \usepackage{bm}
  \usepackage{anysize}
  \usepackage{times,color}
  \usepackage{subfigure}
  \usepackage{mathptmx}
  \usepackage{amssymb}
  \usepackage{amsfonts}
  \usepackage{amsmath}
  \usepackage{pgfplots}

\DeclareGraphicsExtensions{.jpg,.mps,.pdf,.png,.gif}

\begin{document}

\bibliographystyle{plain}

\preprint{APS/123-QED}

\author{L\'{e}onard Monniello}

\affiliation{UPMC Univ Paris 06, UMR 7588, Institut des NanoSciences de Paris, 4 Place Jussieu, F-75005 Paris, France}
\affiliation {CNRS, UMR 7588, Institut des NanoSciences de Paris, 4 Place Jussieu, F-75005 Paris, France}

\author{Antoine Reigue}

\affiliation{UPMC Univ Paris 06, UMR 7588, Institut des NanoSciences de Paris, 4 Place Jussieu, F-75005 Paris, France}
\affiliation {CNRS, UMR 7588, Institut des NanoSciences de Paris, 4 Place Jussieu, F-75005 Paris, France}

\author{Richard Hostein}

\affiliation{UPMC Univ Paris 06, UMR 7588, Institut des NanoSciences de Paris, 4 Place Jussieu, F-75005 Paris, France}
\affiliation {CNRS, UMR 7588, Institut des NanoSciences de Paris, 4 Place Jussieu, F-75005 Paris, France}

\author{\\Aristide Lemaitre}

\affiliation{CNRS, UPR 20, Laboratoire de Photonique et Nanostructures, 
Route de Nozay, F-91460 Marcoussis, France}

\author{Anthony Martinez}

\affiliation{CNRS, UPR 20, Laboratoire de Photonique et Nanostructures, 
Route de Nozay, F-91460 Marcoussis, France}

\author{Roger Grousson}

\affiliation{UPMC Univ Paris 06, UMR 7588, Institut des NanoSciences de Paris, 4 Place Jussieu, F-75005 Paris, France}
\affiliation {CNRS, UMR 7588, Institut des NanoSciences de Paris, 4 Place Jussieu, F-75005 Paris, France}

\author{Valia Voliotis}
\email{voliotis@insp.jussieu.fr}

\affiliation{UPMC Univ Paris 06, UMR 7588, Institut des NanoSciences de Paris, 4 Place Jussieu, F-75005 Paris, France}
\affiliation {CNRS, UMR 7588, Institut des NanoSciences de Paris, 4 Place Jussieu, F-75005 Paris, France}

\title{Non post-selected indistinguishable single photons generated by a quantum dot under resonant excitation}

\begin{abstract}
We report on two-photon interferences from highly indistinguishable single photons emitted by a quantum dot. Stricly resonant excitation with picosecond laser pulses allows coherent state preparation with a significantly increased coherence time ($T_{2}\sim$ 1 ns) and reduced lifetime ($T_{1}\sim$ 650 ps), as compared to a non-resonant excitation scheme. Building-up the Hong-Ou-Mandel dip without post-selection of the interfering photons, visibilities greater than 70 \% have been observed. Near-unity indistinguishable photons could be achieved for every dot if charge noise is controlled. Indeed, the remaining decoherence mechanism is likely due to the fluctuating electrostatic environment of the dots. 
\end{abstract}

\pacs{71.35.-y, 78.55.Cr, 78.67.Hc, 78.47.D-}
\maketitle

Solid state single photon emitters have demonstrated the past decade their high potential for many new applications in the field of nanophotonics \cite{Buckley} and quantum information technology \cite {Imamoglu}. The requirements for efficient on-demand generation of single photons have been partially fulfilled by using quantum dots which have high internal quantum efficiency, embedded in microcavities or photonic crystals for a high extraction efficiency into a specific single mode \cite{QD}. Furthermore, quantum computing schemes with linear optics and quantum teleportation \cite{Knill} require indistinguishable photons, a fundamental property which can be tested by two-photon interferences on a beam splitter in a Hong-Ou-Mandel (HOM) experiment \cite{HOM}. Such two-photon interference experiments have been realized the past few years with post-selection of the single photons emitted by a quantum dot (QD), showing their interesting potential  but also revealing the limitations of incoherent excitation
\cite{HOM_QDs,Patel}. Ideally the photons must be Fourier-transform limited, so the coherence time is trully limited by radiative lifetime. However, for a solid-state emitter like a QD, this is at present difficult to achieve. QDs are strongly interacting with their environment mainly through phonons and trapped charges, leading to dephasing processes \cite{dephase}. Resonant excitation is a necessary condition to keep coherence and a lot of effort has been devoted to this task \cite{resonance}. On resonance excitation allows coherent state preparation increasing the coherence time $T_{2}$ while reducing the spontaneous emission rate $T_{1}$ without the need of a cavity \cite{Enderlin, Monniello}. Moreover, pulsed excitation rather than continuous-wave laser excitation must be used in order to generate single photons in a deterministic way \cite{pulsed, He}. Having on-demand near-unity indistinguishable photons will open the way for realizing entangled photon pairs from one single dot or from remote emitters 
and recent experimental demonstrations are leaning in this direction \cite{entanglement}.

In this Letter, we report on resonant luminescence from a single InAs/GaAs QD under resonant picosecond (ps) pulsed excitation. The QD two-level system, is addressed with $\pi$ pulses corresponding to maximum population on the excited level, a neutral exciton in our case. Independent measurements of lifetime $T_{1}$ and coherence time $T_{2}$ show a degree of indistinguishability $T_{2}/2T_{1}\sim$ 0.7. Second order correlation measurements of the photoluminescence show an antibunching of the order of $g^2(0)=0.07$, with a very low background and without any laser filtering. Two-photon interference on a beam-splitter of two single-photons wave packets without post-selection, show a maximum visibility of 0.73, in very good agreement with the direct measurements of $T_{1}$ and $T_{2}$. Varying the delay between the arrival time of the two photons on the beam-splitter allows to built the Hong-Ou-Mandel dip. The measurements agree very well with the theoretical dependence of the second-order correlation function 
on the time delay without any adjustable parameters.

\begin{figure*}[htb]
\begin{center}
\includegraphics[width=16cm,height=9cm]{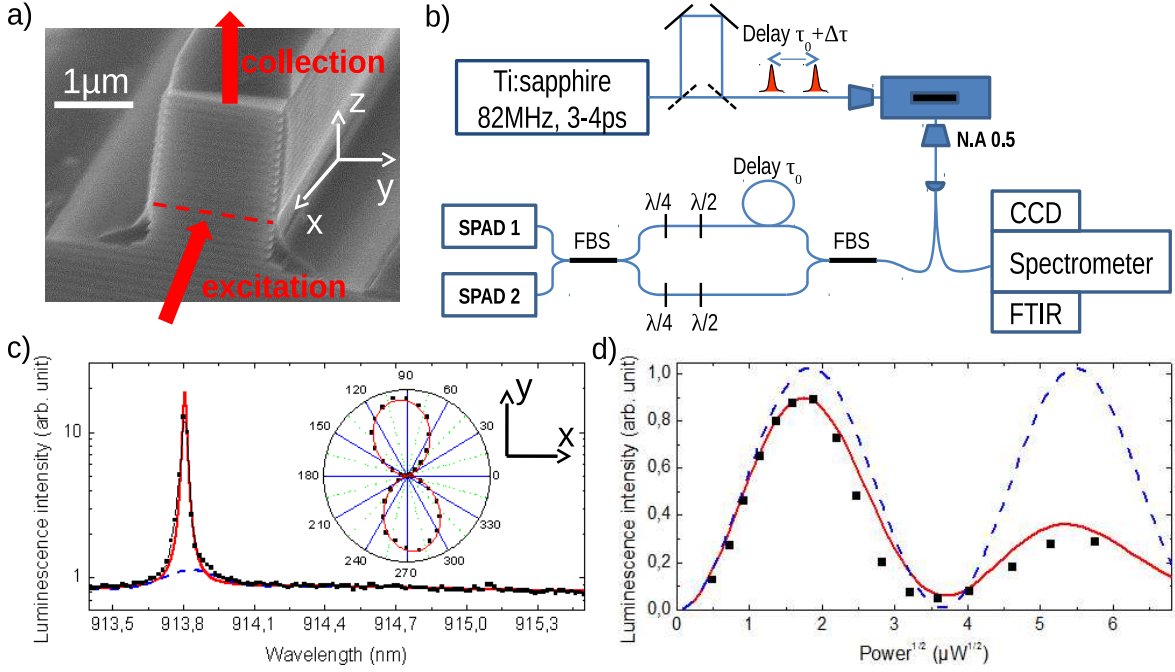}
\end{center}
\caption{(Color online) a) SEM image of one ridge. One can see the Bragg mirrors above and under the QD plane which has been emphasized with a red dashed line. Red arrows show the excitation and the collection paths. b) Schematic drawing of the experimental setup. A pulsed ps Ti:Sapphire laser comes through a first delay line resulting in two pulses separated by $\tau_{0} \pm \Delta \tau$ with $\tau_{0} $ = 3 ns every 12.2 ns. The luminescence is collected by a large N.A. microscope objective, coupled into an optical fiber and sent either into a spectrometer or a fibered Mach-Zender interferometer with two fibered beam splitters (FBS) with fixed $\tau_{0}$ delay for photon correlation studies. A fibered polarization setup equivalent to one $\lambda /2$ and $\lambda /4$ plates controls the outcoming photons polarization. c) Resonant spectrum in semi-logarithmic scale of the studied QD at 7 K : experimental data (black dots) are fitted with a Lorentzian line (red) and a wide gaussian (blue dashed line) 
corresponding to the scattered laser. The 
inset shows the polar diagram of the QD resonant emission. d) Rabi oscillation of the studied QD population. The luminescence intensity is represented by squares and plotted as a function of the square root of the excitation power. The red line is a simulation from optical Bloch equations including excitation-induced damping. The blue dashed line is a simulation of the oscillation without excitation-induced decoherence.}
\label{rabtemp}
\end{figure*}

InAs/GaAs self assembled QDs were grown by MBE on a planar [001] GaAs substrate and embedded in a planar microcavity made of unbalanced AlGaAs/GaAs Bragg mirrors, with 24 pairs below and 12 pairs above the QDs plane (Fig. 1a). The purpose of the Bragg mirrors is just to enhance the luminescence collection efficiency rather than achieving a QD-cavity strong coupling regime with a significant Purcell effect. Therefore the quality factor of the overall structure is low, about 500 but the luminescence signal is increased by a factor 20 to 50. In addition, micrometer ridges (0.8 to 1.2 $\mu$m) are etched on the top surface to design single-mode one dimensional waveguides \cite{Melet}. 
 The QDs are excited along the waveguide by picosecond pulses from a tunable mode-locked Ti:Sapphire laser which polarization along the y axis is imposed by the geometry (Fig. 1a). Thus, on resonance a single eigenstate of the fine structure split exciton state is addressed. In this geometry, the laser is confined in the guided mode enhancing the light-matter interaction and the QD luminescence is collected from the ridge top surface by a confocal microscope (Fig. 1a,b). The scattered laser is thereby greatly suppressed, leading to almost background-free spectra without any need for further polarization filtering. Strictly resonant experiments can then be performed (Fig. 1c). Polar diagrams can be realized to further characterize the QD eigenstates, as shown in the inset in Fig. 1c, where the probed QD emission is linearly polarized, a characteristic of a neutral exciton \cite{Tonin}. The luminescence is coupled into a single-mode optical fiber that can be connected to different setups for spectroscopy, or 
for first order and second order correlation measurements (Fig. 1b). Our setup enables to detect up to 250,000 counts per second on a single-photon avalanche detector (SPAD).

It is worth noticing that resonant excitation is not systematically observed for all the probed QDs. Indeed, it has been reported that resonant excitation can be suppressed due to the presence of trapped charges in the vicinity of the QDs \cite{NGuyen}. In that case, adding a very low power He-Ne laser helps to recover the resonant luminescence. More details on the influence of an additional He-Ne laser will be discussed in the following.
The resonant interaction between the QD and the excitation field gives rise to the well-known Rabi oscillations (RO) of the excited level population. RO are here observed as a function of the pulse area  which is proportional to the square root of the excitation power (Fig. 1d). The damping behavior of such RO under pulsed excitation has been analyzed in Ref [11] for similar samples but without Bragg mirrors. Excitation-induced dephasing processes are mainly due to the resonant coupling between the QD and LA-phonon modes. We have also evidenced that the resonant coupling between the 1D optical mode and the two-level system leads to an acceleration of the radiative lifetime responsible for an additional excitation-induced dephasing mechanism \cite{Domokos}. In the following, all the experiments are perfomed with $\pi$ pulses on-resonance with the neutral exciton (Fig. 1c, d).

As expected, the emission statistics of a QD under resonant excitation corresponds to a single photon source. Second order correlation measurements $g^{(2)}(\tau)$ have been performed (Fig. 2) with 120,000 counts/s on each SPAD with one hour acquisition time. It worths noticing that neither polarization filtering nor spectral selection has been used to suppress the laser scattering and a clear antibunching is shown with a very low multiphoton probability of $g^2(0)=0.07$. The remaining background corresponds exactly to the scattered laser intensity observed in Fig. 1c.

\begin{figure}[t!]
\begin{center}
\includegraphics[width=8cm]{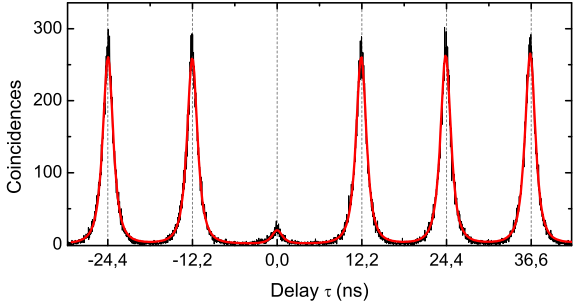}
\end{center}
\caption{(Color online) Second order autocorrelation function $g^2(\tau)$. At zero delay $g^2(0)=0.07$ and is calculated from the ratio of the fitted integrated intensity of the central peak and the six adjacent peaks. The fitting function for each peak is an exponential decay with the exciton radiative lifetime. }
\label{oscrab}
\end{figure}

The on-resonance radiative lifetime $T_1$ has been measured and found to be equal to 670 ps under resonant excitation. For comparison, the lifetime under non-resonant excitation is slightly longer equal to $T_1^{NR}=850$ ps. Normalized data are shown in a semi-logarithmic plot in Fig. 3a, where the time delay between the two curves is due to the radiative cascade that occurs while exciting non-resonantly at high energy \cite{Elvira}. The difference between the two radiative decays is an experimental fact which is systematically observed and can be explained by the strong coupling regime achieved between the QD and the field under resonant excitation. We believe that the resonant coupling modifies the optical density of states thus inducing a modification of the radiative decay rate, similar to an effective Purcell effect \cite{Cohen, Purcell}. 

\begin{figure}[b!]
\begin{center}

	{\includegraphics[width=8cm]{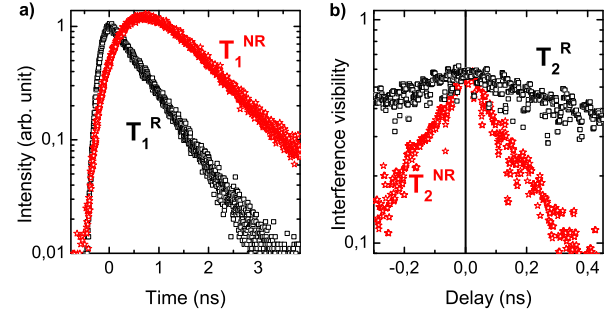}}
\end{center}
\caption{(Color online) a) Lifetime measurements (semi-log scale) for on-resonance (black squares) and non-resonant (red stars) excitation. The temporal decays give $T_1^{R} = 670$ ps and $T_1^{NR} = 850$ ps. b) Fourier transform spectra (semi-log scale) for on-resonance (black squares) and non-resonant (red stars) excitation yielding the coherence time: $T_2^R = 950$ ps and $T_2^{NR}$=200 ps. }
\label{oscrab}
\end{figure}

The coherence time $T_2$ corresponding to the width of the luminescence line has been measured by Fourier transform spectroscopy (Fig. 3b) for resonant and non-resonant excitation. The contrast of the interference fringes as a function of the delay is adjusted by a Voigt profile and gives typical values of $T_2=900$ ps and $T_2=200$ ps for resonant and non-resonant excitation respectively. The inhomogeneous contribution in the Voigt profile is about 10 \%. This result is well understood for non-resonant excitation where electron-hole pairs are photocreated at high energy in the wetting layer and coherence is lost after relaxation into the dot. On the contrary, resonant excitation addresses directly the excited state of the QD and coherence is preserved.

Variations of the resonant coherence time between 850 ps to 950 ps were observed from day to day, as well as variations of the inhomogeneous contribution of the Voigt profile. This effect has been attributed to the fluctuating electrostatic environment \cite{Patel, Houel}. Indeed, the daily thermal cycling of the sample can lead to trapping of charges on defects close to the QDs resulting in a different electrostatic environment seen by the dot . Spectral diffusion can occur and an inhomogeneous broadening of the emission line is observed in this case \cite{Berthelot}. 

A similar behavior has been observed for the radiative lifetime which varies slightly from 650 ps to 700 ps. This effect can also be explained by the local modification of the electrostatic potential that alters the overlap of the hole and electron wavefunctions inside the dot, thus modifying the transition probability. The fluctuating electrostatic environment has also been indirectly observed through a lessening of the resonant luminescence intensity. In that case, the charges may tunnel from a nearby defect into the QD, modifying the first excited state from a neutral to a charged exciton, therefore suppressing the resonance with the excitation laser as reported in Ref. [17]. 
We had also to deal with the same problem and using a very low power (few pW) He-Ne laser, the neutral exciton resonant luminescence intensity was increased tenfold, from 25,000 to 250,000 counts/s on the SPAD. Laser scattering and luminescence due to the He-Ne laser solely, has been estimated to be less than 100 counts/s, comparable to the detector's dark counts.

\begin{figure}[h]
\begin{center}
\includegraphics[width=8cm]{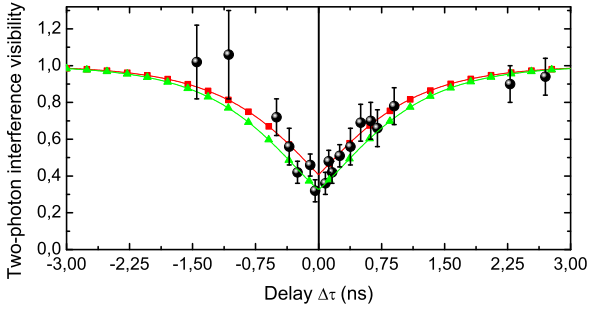}
\end{center}
\caption{(Color online) Two-photon interference experiment showing the Mandel dip. $\Delta\tau$ is relative delay between the two Mach-Zehnder interferometers, counted positively when the excitation interferometer delay is longer than the detection one. The red square and green triangle lines are theoretical evolutions of $g^{(2)}(\Delta\tau)$ for two extreme measured values of $T_1$ and $T_2$ (see text).}
\label{oscrabibi}
\end{figure}

In a two-photon interference experiment, two indistinguishable photons arriving at the same time on a 50/50 beam splitter coalesce and emerge along the same output port of the beam splitter. Then, no simultaneous detection occurs on the two output detectors.
The setup for the realization of the HOM experiment in shown in Fig 1b, where one laser pulse is split into two pulses separated by delay $\tau_{0}=$ 3 ns. The delay line can be adjusted from 1 to 5 ns. After two pulses excitation, the QD emits two sequential photons that are sent into an all-fibered unbalanced Mach-Zehnder interferometer with a fixed delay $\tau_{0}$. We insert in both arms a fibered polarization control setup equivalent to a $\lambda /2$ plate folowed by a $\lambda /4$ in order to compensate the birefringence induced in the optical fibers. Each interferometer output is coupled to an SPAD and for every excitation delay we build the coincidences histogram between the photons arriving on the two detectors, thus extracting the second order correlation function $g^{(2)}(\Delta\tau)$. The data are plotted versus the difference between the two delays in the two interferometers $\Delta \tau$ (Fig. 4). The theoretical dependence of the second-order correlation function on the time delay $\Delta\
tau$, normalized upon the uncorrelated random events, is given by the expression \cite{Bylander}:

\begin{equation}
\begin{split}
g^{(2)}(\Delta\tau) = 1 - \dfrac{2RT}{1-2RT} \Bigl[ \dfrac{T_2}{2T_1} e^{-2|\Delta \tau|/T_2} \\
+ \dfrac{T_2^*}{2T_1} \bigl( e^{-|\Delta \tau|/T_1} - e^{-2|\Delta \tau|/T_2}\bigr) \Bigr]
\end{split}
\end{equation}

where R and T are the (intensity) reflection and transmission coefficients of the beam splitter which has been measured independently to be 60/40 at the QD wavelength. $T_{2}^{*}$ is a pure dephasing time defined by: $\frac{1}{T_2}=\frac{1}{2T_1}+\frac{1}{T_{2}^{*}}$. From the measured values of $T_2$ and $T_1$ we can deduce that $T_{2}^{*}\sim$ 3 ns, which reflects that pure dephasing which is related among others to the presence of fluctuating charges can have an important impact on the coherence properties since it is of the same order of magnitude as $T_2$. The red (squares) and green (triangles) curves in Fig. 4 represent the calculated $g^{(2)}(\Delta\tau)$ using equation (1) with the measured values of $T_2$ and $T_1$. The red (squares) curve correspond to the least advantageous case where $T_2$ is minimum (900 ps) and $T_1$ is maximum (700 ps) while the green (triangles) curve correspond to the most advantageous case with $T_2$ maximum (950 ps) and $T_1$ minimum (650 ps). The experimental data agree 
very well with the calculated curves, except for the longest negative delays where a discrepeancy is observed likely due to experimental uncertainties. Each experimental data point corresponds to one hour acquisition time, which is long enough so fluctuations of the light coupling occur, therefore increasing the noise. The error bars represent the signal to noise ratio for each measurement.

For long delays, the temporal overlap of the two successive photons is reduced and the limit value of 1 is reached as the two photons are totally distinguishable. As $\Delta\tau$ goes to zero, the two photons interfere constructively, until perfect time matching for $\Delta\tau$ = 0 and for totally indistinguishable photons and perfect 50/50 beam splitter, $g^{(2)}(0) = 0$. From equation (1), $g^{(2)}(0) = 1-\frac{2RT}{1-2RT}\frac{T_2}{2T_1}$ and gives a direct value of the degree of indistinguishability defined by the ratio $T_2/2T_1$. In the case of the probed dot, we measure $T_2/2T_1$= 0.73 $\pm$ 0.05, which is also in perfect agreement with the direct measurement of $T_2/2T_1 \sim$ 0.7. 

In summary, we have reported on the first observation of the HOM dip for a neutral QD with strictly resonant pulsed excitation, without polarization post-selection of the emitted photons. The resonant excitation preserves coherence and accelerates the radiative lifetime, enhancing by a factor of 7 the ratio $T_2/2T_1$. Therefore, near-unity indistinguishability of single photons could be reached systematically for every dot once charge noise is controlled. Applying an electric field in a suitably designed structure could be a way to clear out this dephasing mechanism and achieving radiatively-limited optical linewidths.

The authors ackonwledge financial support from the French Agence Nationale de la Recherche (ANR-11-BS10-010) and the C'Nano Ile-de-France (No. 11017728)







\end{document}